\documentclass[twocolumn]{aa}
\usepackage{graphicx}
\usepackage{txfonts}
\usepackage[authoryear]{natbib}
\bibpunct{(}{)}{;}{a}{}{,}
\newcommand{\igr}  {IGR J06074+2205}
\newcommand{\ha}  {H$\alpha$}
\newcommand{\ew}  {EW(H$\alpha$)}

\def\simless{\mathbin{\lower 3pt\hbox
     {$\rlap{\raise 5pt\hbox{$\char'074$}}\mathchar"7218$}}}   
\def\simmore{\mathbin{\lower 3pt\hbox
     {$\rlap{\raise 5pt\hbox{$\char'076$}}\mathchar"7218$}}}   

\def\msun{~{\rm M}_\odot}

\begin{document}

   \title{The optical counterpart to IGR J06074+2205: 
   a Be/X-ray binary showing disc loss and V/R variability
   }

   \subtitle{}
  \author{
	P. Reig\inst{1,2}
	\and
  	A. Zezas\inst{2,3}
         \and
	L. Gkouvelis\inst{2}
          }

\authorrunning{Reig et al.}
\titlerunning{The optical counterpart to \igr}

   \offprints{pau@physics.uoc.gr}

   \institute{IESL, Foundation for Reseach and Technology-Hellas, 71110, 
   		Heraklion, Greece 
	 \and Physics Department, University of Crete, 71003, 
   		Heraklion, Greece 
		\email{pau@physics.uoc.gr}
	 \and Harvard-Smithsonian Center for Astrophysics, 60 Garden
	 Street, Cambridge, MA02138, USA
	}

   \date{Received ; accepted}

\abstract
{Present X-ray missions are regularly discovering new 
X/$\gamma$-ray sources. The identification of the counterparts of these
high-energy sources at other wavelengths is important to determine their
nature. In particular, optical observations are an essential tool in
the study of X-ray binary populations in our Galaxy.
}
{The main goal of this work is to determine the properties of the optical
counterpart to the INTEGRAL source IGR J06074+2205, and study its long-term
optical variability. Although its nature as a high-mass
X-ray binary has been suggested, little is known about its physical
parameters.}
{We have been monitoring IGR J06074+2205 since 2006 in the optical band. We present
optical photometric $BVRI$ and spectroscopic observations covering the
wavelength band 4000-7000 \AA. The blue spectra allow us to determine
the spectral type and luminosity class of the optical companion; 
the red spectra,
together with the photometric magnitudes, were used to derive the
colour excess $E(B-V)$ and estimate the distance.  }
{We have carried out the first detailed optical study of the massive component 
in the high-mass X-ray binary \igr. We find that the optical counterpart to \igr\ is
a $V=12.3$ B0.5Ve star located at a distance of $\sim$4.5 kpc. The monitoring of the
H$\alpha$ line reveals V/R variability and an overall decline of its 
equivalent width. The H$\alpha$ line has been seen to revert
from an emission to an absorption profile. We attribute this variability to 
global changes in the structure of the Be star's circumstellar disc which
eventually led to the complete loss of the disc.
The density perturbation that gives rise to the V/R variability vanishes
when the disc becomes too small.
}
{}


   \maketitle

\begin{table*}
\caption{Photometric measurements of the optical counterpart to \igr.}
\label{phot}\begin{center}
\begin{tabular}{c c c c c c}
\noalign{\smallskip}	\hline\noalign{\smallskip}
Date &  JD (2,400,000+)    &   $B$  &   $V$   &   $R$  & $I$   \\
\noalign{\smallskip}	\hline\noalign{\smallskip}
25 October 2007 &54399.62	&12.88$\pm$0.02  &12.28$\pm$0.03  &11.90$\pm$0.02  &11.47$\pm$0.02    \\
26 October 2007 &54400.29	&12.81$\pm$0.03  &12.26$\pm$0.03  &11.89$\pm$0.04  &11.43$\pm$0.05    \\
\noalign{\smallskip}	\hline
\end{tabular}
\end{center}
\end{table*}
   
\section{Introduction}

IGRJ06074+2205 was discovered by {\em INTEGRAL}/JEM-X during public
observations of the Crab region that took place on 15 and 16 February 2003
\citep{chev04}. The source was detected with a flux of $\sim$ 7 mCrab
($\pm$2 mCrab) in the energy range 3--10 keV and 15 mCrab in the range
10-20 keV range on 15 February 2003. A day later the flux had decreased to
less than 5 mCrab.

Optical spectroscopic observations performed with the MDM  2.4 m telescope
of the brightest stars in the field around the {\em INTEGRAL} position on
27 December 2005 found a Be star within 1' of the {\em INTEGRAL} position,
exhibiting H$\alpha$ in emission with an equivalent width of --6.6 \AA\
\citep{halp05}. Its optical and infrared brightness obtained from the USNO
A2.0 and 2MASS catalogs is $B=13.3$, $R=12.1$, $J=10.49$, $H=10.19$,
$K=9.96$ mag. 

On 2 December 2006 a 5-ks {\it Chandra} observation  improved the accuracy
of the X-ray position and allowed the confirmation of the Be star
suggested by  \citet{halp05} as the correct optical counterpart
\citep{toms06}.   The {\it Chandra} observation also served to discard the
radio source NVSS J060718+220452, which lies 80" from the reported position
of the X-ray source \citep{pool04,pand06}, as a possible radio counterpart
to \igr. {\it Chandra} detected only one source in the {\em INTEGRAL} error
circle at a position of RA = 06h07m26s.62, Dec = 22d05m47s.6 (J2000). The
flux was 2 x $10^{-12}$ erg cm$^{-2}$ s$^{-1}$ in the energy range 0.3-10
keV. The energy spectrum could be fitted with an absorbed power-law with
$N_H= (6\pm2) \times 10^{22}$ cm$^{-2}$ and a photon index  of 1.3 $\pm$
0.8 (90\% confidence errors). The absorbed flux value was nearly 60 times
lower than the value of its discovery, which is consistent with the
behavior of Be X-ray binaries. 

From low-resolution spectroscopic observations obtained using the  G. D.
Cassini 1.5m  telescope in Loiano observatory,  \citet{mase06} suggested a
B8III optical counterpart to \igr. However, their spectra did not have the
spectral resolution to allow a precise classification. In fact, no Galactic
Be/X-ray binary contains a Be star with spectral type later than B3, which
casts some doubt on the accuracy of this classification.

Nevertheless, the available data seem to indicate that \igr\ is a
Be/X-ray binary. Be/X-ray binaries are a class of high-mass X-ray binaries
that consist of a Be star and a neutron star \citep{ziolkowski02}.  The
mass donor in these systems is a relatively massive ($\simmore 10 \msun$)
and fast-rotating ($\sim$80\% of break-up velocity) star, whose equator is
surrounded by a disc formed from photospheric plasma ejected by the star. 
\ha\ in emission  is typically the dominant feature in the spectra of such
stars. In fact, the strength of the Balmer lines in general and of \ha\ in
particular (whether it has ever been in emission) together with a
luminosity class III-V constitute the defining properties of this class of
objects.  Be stars are also observed as single objects, i.e. not forming
part of a binary system \citep{porter03}. The equatorial discs are believed
to be quasi-Keplerian and supported by viscosity \citep{okazaki01}.  The
shape and strength of the spectral emission lines are useful indicators of
the state of the disc. Global disc variations include the transition from a
Be phase, i.e., when the disc is present, to a normal B star phase, i.e.,
when the disc is absent and cyclic V/R changes, i.e., long-term cyclic
changes in the ratio of the blue to red peaks of a split profile that are
attributed to the precession of a density perturbation inside the disc
\citep{okazaki91}.

In this work we present the first detailed study of the optical counterpart
to the X-ray source \igr. Our photometric observations are the first
dedicated observations of the source. So far, the reported magnitudes come
from various catalogs. In most cases it is not possible to know the exact
date of those observations, hence it is not possible to perform variability
studies. In addition, our monitoring of the \ha\ line allows us to
investigate the long-term variability of the circumstellar disc. We confirm
that \igr\ is a Be/X-ray binary. However, its optical companion is more
massive than suggested by \citet{mase06}. The main body of the paper
corresponds to Sect. 3 and 4. In Sect. 3 we present the results of our
spectroscopic observations, with emphasis on the variability of the \ha\
line, while in Sect. 4 we discuss the implications of
the results. Sect. 2 describes the observations and in  Sect. 5 we present
our conclusions.

\begin{figure*}
\resizebox{\hsize}{!}{\includegraphics{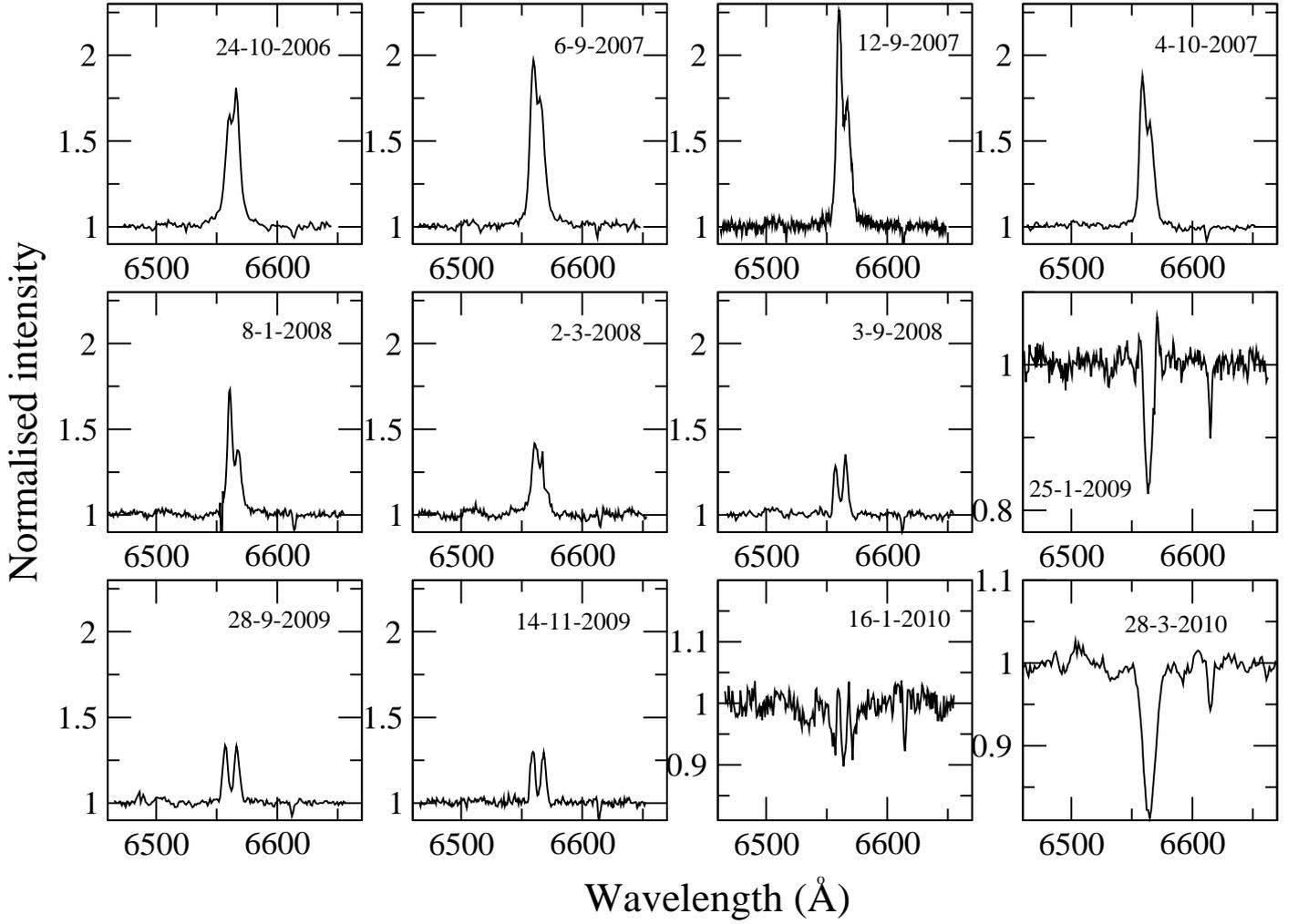}} 
\caption[]{Evolution of the \ha\ line profile. V/R variability as well as a
long-term decrease of the line strength is seen.}
\label{haprof}
\end{figure*}

\begin{table*}
\caption{Log of the spectroscopic observations.}
\label{spec}
\begin{center}
\begin{tabular}{lcllcccc}
\noalign{\smallskip}	\hline \noalign{\smallskip}
Date	&JD (2,400,000+)&Telescope  	&Wavelength	&Resolving	&\ew	&$\Delta_{\rm peak}$	&$\log(V/R)$ \\
	&		&		&coverage (\AA)	&power$^1$	&($\AA$)&(km s$^{-1}$)		&	\\
\noalign{\smallskip}\hline\noalign{\smallskip}
22-10-2006  &54031.58	&SKI	&5845-7200	&2200	&--11.5$\pm$0.4	&--	&--	\\
24-10-2006  &54033.61	&SKI	&5845-7200	&2200	&--11.2$\pm$0.3	&337$\pm$10	&-0.04\\
06-09-2007  &54350.55	&SKI	&5165-7240	&2200	&--12.6$\pm$0.4	&230$\pm$10	&0.05\\
12-09-2007  &54356.98	&FLW	&6050-7050	&7200	&--12.3$\pm$0.5	&275$\pm$5	&0.11\\
04-10-2007  &54378.49	&SKI	&3265-5310	&1200	&--		&--		&--\\
04-10-2007  &54378.53	&SKI	&5165-7240	&2200	&--10.7$\pm$0.5	&303$\pm$13	&0.06\\
25-12-2007  &54460.67	&NOT	&3520-5075	&800	&--		&--	&--\\
08-01-2008  &54474.75	&FLW	&4760-6760	&3200	&--6.8$\pm$0.2	&324$\pm$10	&0.10\\
31-01-2008  &54497.78	&FLW	&4780-6780	&3200	&--6.8$\pm$0.2	&350$\pm$13	&0.08\\
02-03-2008  &54528.65	&FLW	&4770-6770	&3200	&--5.8$\pm$0.5	&334$\pm$25	&0.02  \\
03-09-2008  &54713.56	&SKI	&5090-7165	&2200	&--2.9$\pm$0.1	&370$\pm$6	&-0.02\\
25-01-2009  &54857.68	&FLW	&4775-6775	&3200	&+1.1$\pm$0.1	&595$\pm$25	&-0.01  \\
28-09-2009  &55103.57   &SKI	&5020-7100	&2200	&--3.9$\pm$0.3	&433$\pm$5	&0.0\\
14-11-2009  &55150.90	&FLW	&4755-6760	&3200	&--2.9$\pm$0.2	&407$\pm$5	&0.0 \\
12-01-2010  &55209.87	&FLW	&4730-6735	&3200	&+0.25$\pm$0.15	&--		&0.01  \\
16-01-2010  &55213.71	&FLW	&4740-6740	&3200	&+0.7$\pm$0.2	&525$\pm$25	&0.0  \\
28-03-2010  &55284.85	&NOT	&5785-8285	&1500	&+2.2$\pm$0.2	&--	&--  \\
\noalign{\smallskip}	\hline
\multicolumn{8}{l}{$^1$: at $\sim$6500 \AA\ for the red spectra and at $\sim$4500 \AA\ for the blue spectra} \\
\end{tabular}
\end{center}
\end{table*}

\section{Observations}

Optical spectroscopic and photometric observations of the optical
counterpart to the INTEGRAL source \igr\ were obtained from the 1.3m
telescope of the Skinakas observatory in Crete (Greece) and from the Fred
Lawrence Whipple Observatory at Mt. Hopkins (Arizona). In addition, \igr\
was observed in service time with the Nordic Optical telescope (NOT) and
the  EEV42-40, 2Kx2K chip on the night of 25 December 2007 (Grism\#16) and
on the night 28 March 2010 (Grism\#8). The 1.3\,m telescope of the Skinakas
Observatory (SKI) was equipped with a 2000$\times$800 ISA SITe CCD and a
1302 l~mm$^{-1}$ grating, giving a nominal dispersion of $\sim$1 \AA/pixel.
We also observed \igr\ in queue mode with the 1.5-m telescope (FLW) at Mt.
Hopkins (Arizona), and the FAST-II spectrograph plus FAST3 CCD, a
backside-illuminated 2688x512 UA STA520A chip with 15$\mu$m pixels.  The
observation on 12 September 2007 was obtained with the 1200l/mm grating,
while the rest with the 600l/mm grating. Spectra of comparison lamps were
taken before each exposure in order to account for small variations of the
wavelength calibration during the night. To ensure an homogeneous
processing of the spectra, all of them were normalized with respect to the
local continuum, which  was rectified to unity by employing a spline fit.

The photometric observations were made from the 1.3-m telescope of the
Skinakas Observatory. \igr\ was observed through the Johnson $B$, $V$, $R$ and $I$
filters. For the photometric observations the telescope was equipped with
a  2048$\times$2048 ANDOR CCD with a 13.5 $\mu$m pixel size. Standard stars
from the Landolt list \citep{land09} were used for the transformation
equations.  Reduction of the data was carried out in the standard way using
the IRAF tools for aperture photometry. 

The photometric magnitudes are given in Table~\ref{phot}, while the log of
the spectroscopic observations is shown in Table~\ref{spec}.

\section{Results}

\subsection{The \ha\ line: evolution of spectral parameters}
\label{haevol}

Our monitoring of \igr\ reveals that the \ha\ line is highly variable, both in
strength and shape. The line always shows a double-peaked profiles but the
relative intensity of the blue (V) over the red (R) peaks varies.

Table~\ref{spec} gives the log of the spectroscopic observations and some
important parameters that resulted from fitting two Gaussians to the \ha\ 
line profile. Column 6 gives the equivalent width of the \ha\ line
(EW(H$\alpha$)). The main source of uncertainty in the equivalent width
stems from the always difficult definition of the continuum. The \ew\ given
in Table~\ref{spec} correspond to the average of twelve measurements, each
one from a different definition of the continuum and the error
is the scatter (standard deviation) present in those twelve measurements. 

Column 7 gives the peak separation, $\Delta_{\rm peak}$, of the blue and
red peaks. The peak separation is simply the difference between the central
wavelength of the red minus the blue peak in velocity units ($\Delta
\lambda/\lambda \times c$), where $c$ is the speed of light.   The
errors are estimated by propagating the uncertainty in the determination of
the central wavelength of the gaussian profile used to fit the \ha\
profile. The uncertainty of the best-fit peak wavelengths are at
68\% confidence intervals. The peak separation is related to the size of
the Be star's disc since 

\begin{equation} 
\frac{R_{\rm disc}}{R_*}=\left(\frac{2 v \sin i}{\Delta_{\rm peak}}\right)^2 
\end{equation}

\noindent where $v\sin i$ is the projected rotational velocity of the B
star ($v$ is the equatorial rotational velocity and $i$ the inclination
of the equatorial plane with respect to the observer).

Column 8 shows the ratio between the core intensity of the blue and red
humps. The V/R ratio is computed as the logarithm of the ratio of the
relative fluxes  at the blue and red emission peak maxima (without prior
correction for the underlying continuum level). Thus negative values
indicate a red-dominated peak, that is, $V<R$.

\begin{figure}
\resizebox{\hsize}{!}{\includegraphics{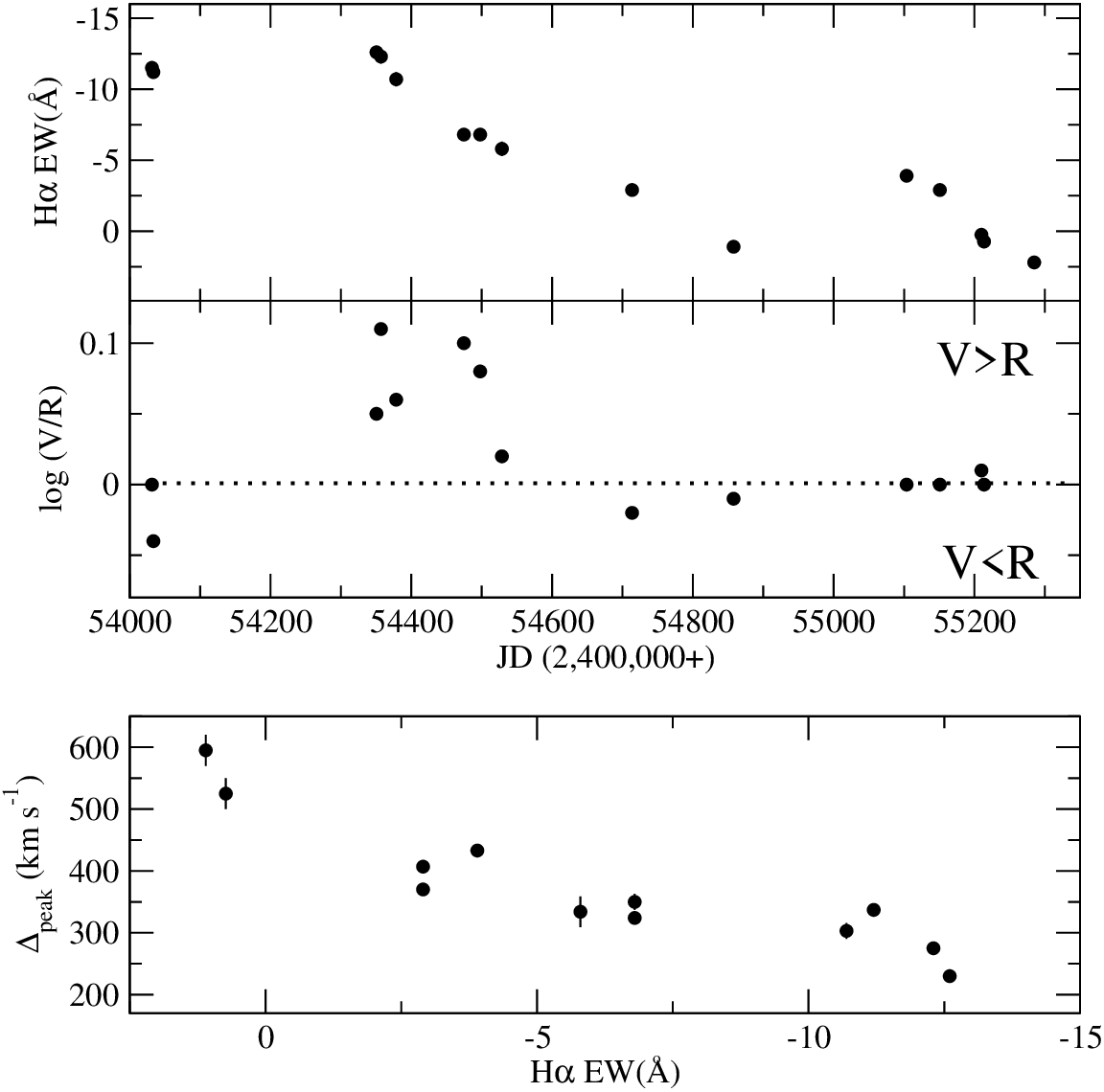}} 
\caption[]{Evolution of the \ha\ equivalent width (top) and V/R ratio (middle)
with time.  
The bottom panel shows the peak separation as a function of EW(H$\alpha$).}
\label{specpar}
\end{figure}

Figure \ref{haprof} displays the evolution of the line profiles. V/R
variability is clearly seen, indicating significant changes in the
structure of the equatorial disc on timescales of months. In addition, a
long-term weakening of the disc is suggested by the decrease of the
equivalent width (Fig.~\ref{specpar}), which changed from $\sim$--12 \AA\
in September 2007 to $\sim$--3 \AA\ a year later. Negative values of the
equivalent width mean that the line displays an emission profile. In 2010,
\ew\ was positive, indicating an absorption profile.

\subsection{Spectral classification}

\begin{figure*}
\begin{center}
\includegraphics[width=16cm,height=10cm]{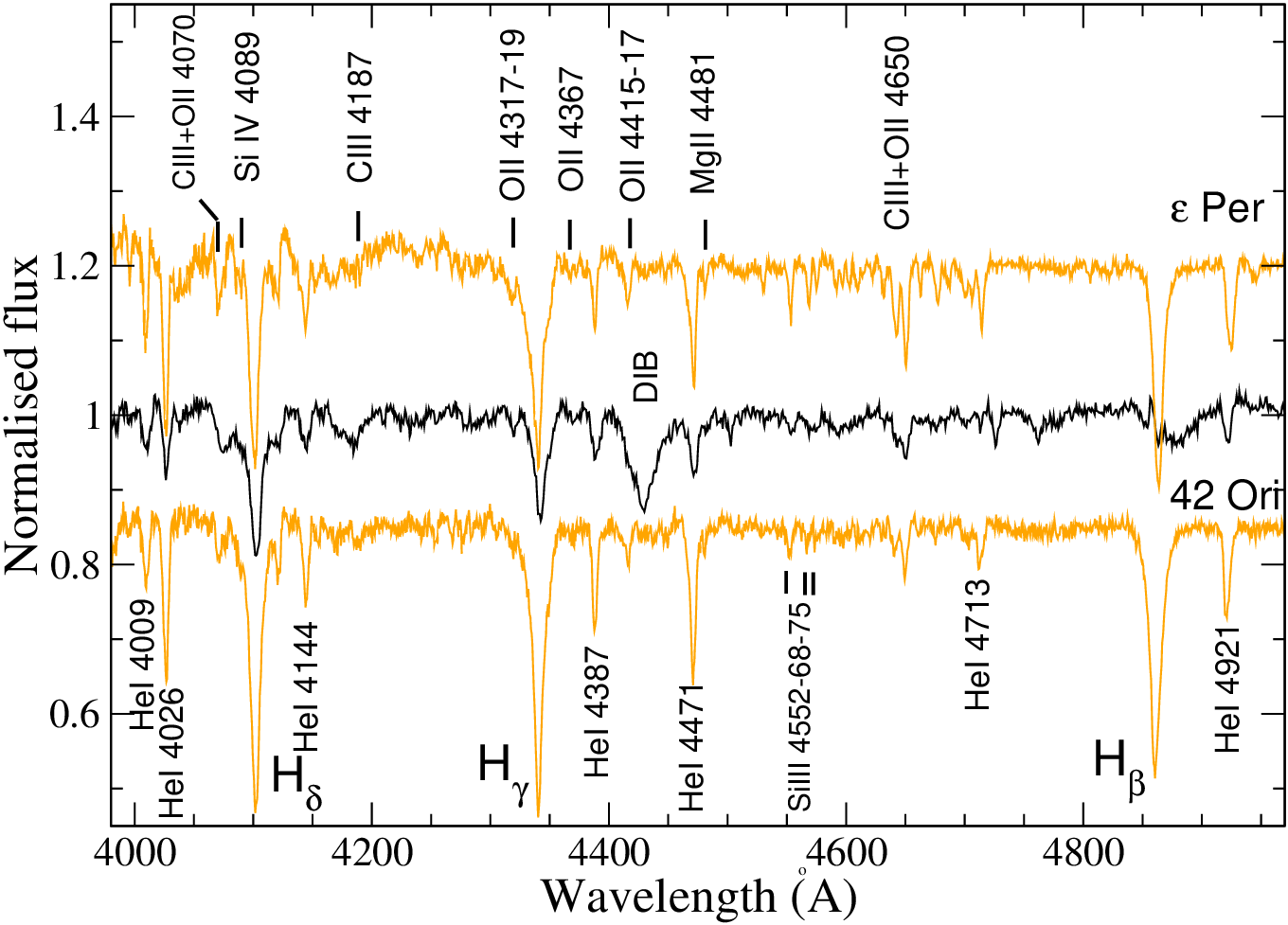} 
\caption[]{Identification of the He and metallic lines used for spectral
classification in \igr (middle spectrum). A comparison with two MK standards 
is provided: the B0V $\epsilon$ Per (top) and the B1V 42 Ori (bottom).}
\label{speclass}
\end{center}
\end{figure*}

The blue spectrum of \igr\ (Fig.~\ref{speclass}, middle) is dominated by
hydrogen and neutral helium absorption lines, clearly indicating an
early-type star (O or B). The Balmer series lines from H$\beta$ up to
H$\eta$ at 3835 \AA\ are seen in absorption. No or very weak He II lines
($\lambda$4541, $\lambda$4686) are present, which implies a type later than
B0. The weak Mg II $\lambda4481$ indicates a type earlier than B2. The
strong  C III + O II blend allows us to constrain better the upper limit as
its presence indicates a type earlier than B1.5.  The primary
classification criteria in the spectral range O9-B1.5, namely the ratio of
Si IV at $\lambda$4686 to Si III at $\lambda$4552-68-75, cannot be used
because the fast rotation blurs the weak Si IV line. Nevertheless, Si III
4552-68-75 is clearly present, which if the star is on the main sequence,
it would imply a spectral type earlier than B2; Si III has its
maximum strength at type B0.5V and is no longer seen at B2V \citep{walb90}. 

As for the luminosity class, the relative strength of the Si
III$\lambda$4552-68-75 complex and the CIII + OII blends compared to that
of nearby He I lines favours a main sequence classification.  Likewise, the
low relative intensity  of OII lines also indicates a luminosity class V
star. 

A visual comparison of the \igr\ spectrum to those of the standards shown
in \citet{walb90} yields that the spectrum of HD 36960 as the most similar
one, hence we assign a spectral type B0.5V to \igr. However a slightly
later type of B1V cannot be ruled out. Fig.~\ref{speclass} also shows the
spectra of two MK standard stars, namely, $\epsilon$ Per (a B0V star) and 42
Ori (a B1V star).


\subsection{Reddening and distance}

To estimate the distance, the amount of interstellar extinction $A_{\rm
V}=R \times E(B-V)$ to the source has to be determined.  The observed colour of
\igr\ is $(B-V)=0.58\pm0.03$, while the expected one for a B0.5V star
$(B-V)_0=-0.26$ \citep{john66,guti79,wegn94}. Thus we derive a colour
excess of $E(B-V)=0.84\pm0.03$. Assuming the standard extinction law
$R=3.1$ and an average absolute magnitude for a B0.5V star of
$M_V=-3.5\pm0.5$ \citep{vacc96,wegn06} the distance to \igr\ is estimated
to be  4.4$\pm$1.0 kpc. Since the photometric magnitudes were
obtained when the source was displaying a relatively large value of the
\ew, this estimate of the distance should be taken as a lower limit. In the
absence of the disc the colour $B-V$ is expected to be smaller (more
"blue"), which would imply a lower value of $E(B-V)$, hence longer
distance.

The colour excess can also be estimated from the strength of the diffuse
interstellar bands \citep{herb75,herb91,gala00}. Using the strongest lines
(6203\AA, 6269\AA, 6276-79\AA, and 6613\AA) in the red spectrum of
\igr, the estimated color excess is E(B-V)=0.85$\pm$0.12, in excellent
agreement with the photometric derived value, hence giving support to the
spectral classification suggested.  The error is the standard
deviation of the measurements of all employed lines.

\subsection{Rotational velocity}

Be stars are fast rotators. They have, on average, larger observed
rotational velocities than B stars as a group \citep{slet82}. The
determination of the rotational velocity is important because it is
believed to be a crucial parameter in the formation of the circumstellar
disc. A rotational velocity close to the break-up or critical velocity
(i.e. the velocity at which centrifugal forces balance  Newtonian gravity)
reduces the effective equatorial gravity to the extent that weak
processes such as gas pressure and/or non-radial pulsations may trigger
the ejection of photospheric matter with sufficient energy and angular
momentum to make it spin up into a Keplerian disc. At present there is no
consensus on how close to their critical velocity Be stars rotate, although
observations suggest that a large fraction of Be stars rotate at 70--80\% of
the critical value \citep{slet82,porter96,yudi01}. 

The projected rotational velocity can be estimated by measuring the full
width at half maximum (FWHM) of \ion{HeI}\ lines \citep{stee99}. The
projected rotational velocity was obtained as the average of the values
from three \ion{HeI}\ lines at 4026 \AA, 4143 \AA, and 4471 \AA. We
measured the width of these lines from the NOT spectrum as it provides the
highest resolution in our sample. We made three different selections of the
continuum and fitted Gaussian profiles to these lines.  We obtained $v \sin
i=260\pm20$ km s$^{-1}$. The value quoted is the weighted average of the
nine measurements.

Table~\ref{rotvel} gives the rotational velocity of some Be/X-ray binaries
obtained from the reference in column 8.
Column 6 gives the inclination angle of the orbit with respect to the line
of sight, which allows to estimate the true rotation velocity of the Be
star. The critical velocity depends on the spectral type. For Be/X-ray
binaries whose spectral type distribution spans a very narrow range
(O9-B2), the critical or break-up velocity is 500-600 km s$^{-1}$, with the
lower end corresponding to later spectral types \citep{cran05}. The term
{\em shell} refers to double-peak lines whose central depression  is lower
than the stellar continuum \citep[see e.g.][]{humm00}. This type of profile
occurs in systems with high inclination angles, i.e., the so called edge-on
systems.

\begin{table*}
\begin{center}
\caption{Comparison of \igr\ with other Be/X-ray binaries. }
\label{rotvel}
\begin{tabular}{lllccccc}
\noalign{\smallskip} \hline \noalign{\smallskip}
X-ray		&Optical	&Spectral&Disc-loss	&P$_{\rm orb}$	&Inclination	&$v \sin i$	&Reference \\
source		&counterpart	&type	&episodes	&(days)		&angle ($^{\circ}$)&(km s$^{-1}$)& \\
\noalign{\smallskip} \hline \noalign{\smallskip}
{\em \igr}	&--		&B0.5IV	&yes	&--	&--	&260$\pm$20	&this work \\
4U 0115+634	&V635 Cas	&B0.2V	&yes	&24.3	&43	&300$\pm$50	&1 \\
RX J0146.9+6121	&LS I +61 235	&B1III-V&no	&--	&--	&200$\pm$30	&2 \\
V 0332+53	&BQ Cam		&O8-9V	&no	&34.2	&$<$10	&$<$150		&3 \\
X-Per		&HD 24534	&O9.5III&yes	&250	&23-30	&215$\pm$10	&4,5  \\
RX J0440.9+4431	&LS V +44 17	&B1III-V&yes	&--	&--	&235$\pm$15	&6 \\
1A 0535+262	&HD 245770 	&O9.7III&yes	&111	&28-35	&225$\pm$10	&7,8 \\
RX J0812.4-3114	&LS 992		&B0.5III-V&yes	&81.3	&--	&240$\pm$20	&9 \\
1A 1118-615	&Hen 3-640	&O9.5IV	&no	&--	&--	&300		&10 \\
4U 1145-619	&V801 Cen 	&B0.2III&no	&187	&$<$45	&250$\pm$30, 290&11,12  \\
4U 1258-61	&V850 Cen 	&B2V	&yes	&132	&90 shell&$<$600	&13 \\
SAX J2103.5+4545&--		&B0V	&yes	&12.7	&--	&240$\pm$20	&14 \\
\noalign{\smallskip} \hline
\multicolumn{8}{l}{[1] \citet{negu01}, [2] \citet{reig97}, [3] \citet{negu99}, [4] \citet{lyub97}}\\
\multicolumn{8}{l}{[5] \citet{delg01}, [6] \citet{reig05}, [7] \citet{haig04}, [8] \citet{grun07}}\\
\multicolumn{8}{l}{[9] \citet{reig01}, [10] \citet{jano81}, [11] \citet{jano82}, [12] \citet{webs74}}\\
\multicolumn{8}{l}{[13] \citet{park80}, [14] \citet{reig04}}\\
\end{tabular}
\end{center}
\end{table*}

\section{Discussion}

All spectroscopically identified optical companions of  Be/X-ray binaries
in the Milky Way have spectral type earlier than B3. To the authors'
knowledge there is no one single exception. In terms of the mass, this
means that there are no Be stars in Be/X binaries with masses lower than 8
$\msun$.  Thus the classification of \igr\ as a B8III reported by
\citet{mase06} was surprising.

This narrow range of spectral types in Be stars which are part of binary
systems contrast with the wide range found in isolated systems
\citep{negu98}. The "Be phenomenon" can be observed from late O stars to
early A stars. Such narrow range in masses for Be/X-ray binaries can be
explained by invoking a non-conservative binary evolution. \citet{port95}
showed that if during the evolution of the binary the amount of angular
momentum lost per unit mass through the second Lagrangian point increases
then the number of late-type Be stars with a neutron star companion
reduces. The reason for this is that a significant loss of angular momentum
implies that a larger number of binaries ends up as mergers.  On
similar grounds, \citet{bever97} compared two models that only differed in
the amount of angular momentum lost during non-conservative Roche Lobe
overflow and found that the minumum mass of the mass-gainer star
corresponding to the model with the higher loss of angular momentum was
$\sim 7 \msun$. The spectral type that we determine for the optical
counterpart of \igr, namely B0.5V is in good agreement with the spectral
type distribution of galactic Be/X-ray binaries. 

The \ha\ line shows V/R variability, that is, the cyclic variation of the
relative intensity of the blue (V) and red (R) peaks in the split profile
of the line. The V/R variability is believed to be caused by the gradual
change of the amount of the emitting gas approaching the observer and that
receding from the observer due to the precession of a density perturbation
in the disc. Double-peak symmetric profiles are expected when the
high-density part is behind or in front of the star, while asymmetric
profiles are seen when the high-density perturbation is on one side of the
disc. More precisely, when the high-density part of the disc is moving
toward the observer we expect to see a blue-dominated $V>R$ profile, while
when the high-density part is receding from the observer red-dominated
profiles $V<R$ are expected \citep{telt94}. For systems with high
inclination angles, the two symmetric cases can be readily distinguished
since the central depression between the two symmetric peaks is much more
pronounced (reaching or going beyond the continuum), adopting a shell
profile, when the perturbation is behind the star. If the density
perturbation revolves around the star in the same direction as the material
in the disc (prograde precession) then  the V/R sequence would be
\citep{telt94}: $V=R$ (perturbation behind the star) $\longrightarrow$
$V>R$ $\longrightarrow$ $V=R$ (perturbation in front of the star, shell
profile) $\longrightarrow$ $V<R$.

Figure \ref{haprof} shows the \ha\ line profile from most of the spectra that
we obtained. The first recorded spectrum displays a red-dominated peak,
$V<R$. The relative intensity of the two peaks appeared reversed, i.e.
$V>R$, about a year later. If we interpreted the January 2009 line as an
emission line affected by a strong shell component (self-absorption from
the disc), then  the  $V>R$ would be followed by a shell phase and we would
conclude that the motion is prograde. 

The available data do not allow us to constrain the duration of the V/R
cycle because a major structural change occurred which brought the V/R
cycle to an end. This major event was the loss of the equatorial disc. 
Nevertheless, we can constrain the quasiperiod of the V/R cycle from the
two complete phases covered by the data, namely, the $V>R$ and shell
phases. The shell phase would have lasted for about 14-16 months
(July-August 2008 to November -December 2009). Due to the observational
gaps the $V>R$ phase is less constrained. The maximum duration for this
phase is approximately 20 months (December 2006 to August 2008). Since we do
not expect the transition between phases to be abrupt, it reasonable to
limit the $V>R$ phase to 15-17 months (February 2007 to June 2008).
Assuming then that each V/R phase lasts for about 15 months the V/R
quasiperiod is estimated to be $\sim$ 5 years. 

The long-term decline of the H$\alpha$ equivalent width
(Fig.~\ref{specpar}) undoubtedly implies a progressive weakening of the
disc over a period of $\sim$3 years. Moreover, the latest spectrum in March
2010 displays an absorption profile. The reversion from emission to
absorption is generally believed to be due to the disappearance of the
disc. Nevertheless the fact that the line still shows some residual
emission at the wings and that, although positive, the \ha\ equivalent
width, +2.2, is still somehow smaller than expected for a non-emitting
B0.5V star, which according to \citet{jasc87} should be $\sim$ +3.5-4 \AA\
may imply that the disc-loss phase continues at the time of writing. 

Irrespectively of whether the disc vanished completely or not, the data
show one important result, namely, the V/R variability is absent in small
discs. In other words, only in well developed discs can the density
perturbation survive. We have searched the literature for spectral line
variability immediately before and after disc-loss episodes in other
Be/X-ray binaries (see column 4 in Table~\ref{rotvel}).  We observe that
shortly before the diappearance of the disc and during the first instances of
the formation of a new disc the \ha\ line profile is symmetric and double
peaked.  None of the Be/X-ray binaries that have gone through disc-loss
phases and for which there is a good optical follow-up coverage, namely, X
Per in 1990 \citep{clar01}, 4U 0115+63 in 1997 \citep{negu01} and in 2002
\citep{reig07}, 1A 0535+262 in 1999 \citep{haig99,grun07} and RX J0440.9+4431 in
2001 \citep{reig04} exhibited asymmetric profiles during the  initial
stages of disc formation. That is, the effects of the density perturbation
do not show up until the disc is fully developed.  The same result holds
for the final stages of disc loss. Just before the detection of the
absorption profile the V/R ratio is $\sim$ 1.

Using the equation given in Sect.~\ref{haevol}, the rotational velocities
given in Table~\ref{rotvel} and the peak separation of the first available
asymmetric profile after the disc loss $\Delta_{\rm peak}=220$ km s$^{-1}$
\citep[X-Per,][]{clar01}, $\Delta_{\rm peak}=305$ km s$^{-1}$ \citep[4U
0115+63,][]{negu01} we find that the disc must reach a radius of $\sim 4
R_*$ for the onset of the V/R variability. Virtually the same value is
derived from Fig. 5 in \citet{grun07} for 1A 0535+262. For RX J0440.9+4431
\citep{reig04} only a lower limit, $>2 R_*$, can be given. 

The data {\em before} the disc-loss phase is even more scarce. Nevertheless,
the available data seems to indicate that, before the disc loss, the
density wave may survive in smaller discs ($\sim 1.5-2 R_*$). In 4U
0115+63, the last available spectrum prior to the disc-loss event showing
substantial V/R variability corresponds to a disc radius of 1.6 $R_*$. In
\igr, the peak separation measured in the spectrum taken  on 1 January 2008
corresponds to radius of 2.6 $R_*$. 

Note that the peak separation does not necessarily indicate the
geometrical size of the disk, but rather the radial distance at which \ha\
emission becomes optically thick. How well this correlates with the disc
size depends on its density distribution. It is generally assumed that the
density falls off as a power law with increasing distance from the star, 
$\rho(r)=\rho_0\left(\frac{r}{R_*}\right)^{-n}$. The fact that  the
power-law index $n$ estimated observationally lies in a very narrow range
$3 < n < 3.5$  \citep{waters88,jones08} and that the relevant radius in
this order of magnitude calculation is that of the \ha\ emitting
region justifies the validity of the comparison.

Given the large observational gaps in the present observations it is clear that
further long-term spectroscopic monitoring of these and other systems are
required to confirm these results and set better constraints on the
variability timescales and behaviour of the disc perturbations.

\section{Conclusion}

We have performed optical photometric and spectroscopic observations of the
optical counterpart to \igr. From the photometric magnitudes and colours
and the strength of various diffuse interstellar bands we have estimated
the distance to be $\sim$ 4.5 kpc. From the ratios of various metallic lines
we have derived a spectral type B0.5V and from the width of three He I lines
estimated the rotation velocity of the underlying B star in 260 km
s$^{-1}$.

The long-term optical variability of this system is characterised by global
changes in the structure of the equatorial disc around the Be star
companion. These global changes manifest as asymmetric profiles of the \ha\
line and a significant decay of its intensity.  The asymmetry consists of
V/R variability with characteristic timescales of the order of few months.
The  \ha\ line has been seen to change from emission to absorption,
indicating the loss of the equatorial disc. The V/R cycle stopped when the
equatorial disc became too small. A small disc cannot support a density
perturbation. During the final stages of the disc evolution V/R asymmetries
can be seen up to a disc radius of $\sim$2 $R_*$. The density wave fades away
before the complete dissipation of the disc because the disc became too
tenuous to support the density perturbation. A comparison with other
systems reveal that, after the loss of the disc, a density perturbation
does not develop until the disc radius is of the order of 4 $R_*$.

\begin{acknowledgements}

We thank the observers P. Berlind and M. Calkins for performing the FLWO
observations. This work has been  supported in part by the European Union
Marie Curie grant MTKD-CT-2006-039965 and EU FP7 "Capacities" GA No206469. 
This work has made use of NASA's Astrophysics Data System Bibliographic
Services and of the SIMBAD database, operated at the CDS, Strasbourg,
France. Skinakas Observatory is a collaborative project of the University
of Crete, the Foundation for Research and Technology-Hellas and the
Max-Planck-Institut f\"ur Extraterrestrische Physik.

\end{acknowledgements}

\end{document}